\documentclass[]{pasj02} 
\usepackage[switch,mathlines]{lineno} 
\usepackage{natbib} 
\usepackage{color}

\DeclareAbbreviation\jcap{JCAP}
\DeclareAbbreviation\pasa{PASA}

\jyear{2024}
\Received{}
\Accepted{}

\begin{document} 

\title{Stacked strong and weak lensing united: Improved measurement of the stellar and dark matter distributions in massive early-type galaxies at $z\sim 0.5$}

\author{
  Momoka \textsc{Fujikawa}\altaffilmark{1}\altemailmark\email{25wm2122@student.gs.chiba-u.jp}
  and 
 Masamune \textsc{Oguri}\altaffilmark{1,2}\altemailmark\orcid{0000-0003-3484-399X}\email{masamune.oguri@chiba-u.jp} 
}
\altaffiltext{1}{Department of Physics, Graduate School of Science, Chiba University, 1-33 Yayoi-Cho, Inage-Ku, Chiba 263-8522, Japan}
\altaffiltext{2}{Center for Frontier Science, Chiba University, 1-33 Yayoi-cho, Inage-ku, Chiba 263-8522, Japan}



\maketitle

\begin{abstract}
We present a new measurement of the stellar and dark matter distributions in massive early-type galaxies at redshift $0.4<z<0.6$ by combining stacked strong and weak lensing measurements. The stacked weak lensing measurements down to the small radius of $0.013\,\mathrm{Mpc}/h$ enabled by the Subaru Telescope Hyper Suprime-Cam data are combined with stacked enclosed projected mass measurements from 13 strong lens systems. We find that adding strong lensing constraints improves constraints on the stellar and dark matter distributions. The central dark matter profile is found to be consistent with the standard Navarro-Frenk-White density profile. The stellar initial mass function is found to be bottom-heavy. The total density profile in the inner region is  described well by the power-law density profile with the slope of $-1.96\pm 0.07$, which can also explain weak lensing signals out to $1\,\mathrm{Mpc}/h$ reasonably well. The stacked weak lensing profile constrains an internal mass-sheet transformation parameter to $\lambda_{\mathrm{c}}=0.978\pm0.010$ {assuming a single cored component}. Our analysis demonstrates that the combination of the stacked strong and weak lensing serves as a powerful tool for studying central density profiles of galaxies.
\end{abstract}


\section{Introduction}

Precise measurements of dark matter distributions in galaxies serve as a useful test of the standard cold dark matter (CDM) model \citep[e.g.,][for reviews]{2017ARA&A..55..343B,2026arXiv260728564N}. Such observational tests are important particularly because the standard CDM model faces a number of challenges at small scales, including the so-called core-cusp problem, which is the mismatch of central density structures of dark matter halos between numerical simulations assuming the standard CDM model and observations mainly in dwarf galaxies. This mismatch can be explained by either the baryonic feedback effect or the modification of the nature of dark matter. To clarify its origin, it is crucial to measure dark matter distributions for various types of galaxies with different masses and redshifts.

In order to measure central dark matter distributions in galaxies accurately, one has to disentangle them from stellar mass distributions that usually dominate at centers of galaxies. Furthermore, direct measurements of stellar masses of galaxies have their own importance because they enable one to constrain the stellar initial mass function (IMF) that is a key element connecting the small-scale star formation to the galaxy evolution and the chemical enrichment \citep[e.g.,][for reviews]{2010ARA&A..48..339B,2018PASA...35...39H,2020ARA&A..58..577S}. Given observational indications for the variation of the stellar IMF across different mass and redshift, it is again crucial to measure the stellar IMF for various types of galaxies.

Focusing on early-type galaxies, their central mass distributions have been studied mainly using stellar dynamics and strong gravitational lensing. These observations reveal that the total density profile is roughly consistent with a singular isothermal profile with $\rho(r)\propto r^{-2}$ \citep[e.g.,][]{2006ApJ...649..599K,2007ApJ...667..176G,2009ApJ...703L..51K,2011MNRAS.415.2215B,2012ApJ...757...82B,2013MNRAS.432.1709C,2013ApJ...777...98S,2018MNRAS.480..431L,2019MNRAS.489.5483T,2023MNRAS.521.6005E}. Disentangling the stellar and dark matter distributions is also attempted using stellar dynamics \citep[e.g.,][]{2012Natur.484..485C,2016ARA&A..54..597C,2017MNRAS.468.3949A,2022ApJ...930..153S,2023A&A...672A..84D,2024MNRAS.527..706Z,2024MNRAS.528.5295Y} and strong gravitational lensing \citep[e.g.,][]{2003ApJ...595...29R,2010ApJ...709.1195T,2010ApJ...721L.163A,2012ApJ...747L..15G,2014MNRAS.439.2494O,2019A&A...630A..71S,2025A&A...697A..95S,2026MNRAS.548ag683L}. Strong lensing and stellar dynamics are often combined to obtain tighter constraints \citep[e.g.,][]{2004ApJ...611..739T,2010ApJ...721L...1T,2015ApJ...800...94S,2015ApJ...814...26N,2018MNRAS.476..133O,2021MNRAS.503.2380S,2025MNRAS.541....1S,2026arXiv260326894W}.

Additional constraints on mass distributions of early-type galaxies can be obtained by weak gravitational lensing. \citet{2007ApJ...667..176G} measure a stacked weak lensing signal around a Sloan Lens ACS (SLACS) strong lens sample to find that the total mass profile is nearly isothermal out to $\sim 300\,\mathrm{kpc}/h$. {This weak lensing measurement pipeline is used in \citet{2021MNRAS.503.2380S} to study the dark matter halo of the SLACS strong lens sample from a joint analysis of strong lensing, stellar dynamics, and weak lensing.}  \citet{2018MNRAS.481..164S} combine constraints from the SLACS strong lens sample with weak lensing measurements of a controlled galaxy sample using the Hyper Suprime-Cam Subaru Strategic Program \citep[HSC-SSP;][]{2018PASJ...70S...4A} data to discuss the variation of the mass-to-light ratio of massive galaxies. {\citet{2024ApJ...973..102X} measures the total mass profile of massive galaxies down to $\sim 20\,\mathrm{kpc}$ using stacked weak lensing magnifications and finds an excess compared with simulations.}
\citet{2024MNRAS.533..795K} propose to use the stacked weak lensing analysis of massive galaxies to break the mass-sheet degeneracy in strong lens mass modeling.

In our previous paper \citep{2026OJAp....961580F}, we further pursue this direction and constrain the stellar and dark matter distributions in early-type galaxies at $z\sim 0.5$ directly with the stacked weak lensing analysis using the HSC-SSP data. We demonstrate that the high-quality HSC-SSP weak lensing data enable us to measure stacked weak lensing signals down to $\sim 10\,\mathrm{kpc}/h$, from which we find non-zero core radii of dark matter distributions in galaxies with stellar masses of $\sim 10^{11}M_\odot$. Additionally, our direct stellar mass measurement with the stacked weak lensing analysis suggests the bottom-heavy stellar IMF.

In this paper, we combine stacked weak lensing measurements presented in \citet{2026OJAp....961580F} with strong lensing measurements of Einstein radii to improve constraints on the stellar and dark matter distributions for galaxies in the highest stellar mass bin $10^{11.5}M_\odot<M_{\star,\mathrm{in}}<10^{11.7}M_\odot$ studied in \citet{2026OJAp....961580F}. For this stellar mass bin, the stacked weak lensing measurement is found to be relatively noisy due to the small number of galaxies in the stellar mass bin. We show that adding strong lensing constraints indeed significantly improves the constraints. We also study the total mass profile of galaxies using both the strong and weak lensing constraints. While our approach to combine strong and weak lensing measurements is partly similar to \citet{2018MNRAS.481..164S}, there are some notable differences between them, including different models of dark matter distributions and different lens redshifts.

The structure of this paper is as follows. In Section~\ref{sec:data}, we describe strong and weak lensing data used for the analysis, as well as our method to fit the data. In Section~\ref{sec:smdm}, we present our analysis results of stellar and dark matter distributions. In Section~\ref{sec:total} we analyze the total density profile using both strong and weak lensing data. We give our conclusion in Section~\ref{sec:conlusion}. Throughout the paper, we assume a flat Universe with the matter density parameter $\Omega_{\mathrm{m}}=0.3$, the cosmological constant $\Omega_{\Lambda}=0.7$, the baryon matter density parameter $\Omega_{\mathrm{b}}=0.05$, and the dimensionless Hubble constant $h=0.7$.

\section{Data and method}\label{sec:data}


\subsection{Weak lensing data}
Details of the weak lensing analysis are presented in \citet{2026OJAp....961580F}, and here we provide a brief summary of the data used for the weak lensing analysis.

We use the HSC-SSP three-year shear catalog
\citep{2022PASJ...74..421L} for our weak lensing analysis. This
catalog covers an area of $\sim 430$~deg$^2$ with a raw galaxy
number density of $\sim 23$~arcmin$^{-2}$, and it is validated with a
series of null tests for systematics. For a foreground lens sample, we employ the HSC-SSP final-year
photometric luminous red galaxy (LRG) sample \citep{2026PASJ...78..416O}.
The photometric LRG sample is constructed based on the stellar
population synthesis fitting of multi-band magnitudes of individual galaxies with the model of quiescent early-type galaxies based on \citet{2003MNRAS.344.1000B}, which is also used for the CAMIRA cluster finding algorithm
\citep{2014MNRAS.444..147O,2018PASJ...70S..20O,2026PASJ...78..416O}.
The final-year photometric LRG sample covers an area of $\sim
1200$~deg$^2$ and contains galaxies in the photometric redshift range
of $0.05\leq z_{\mathrm{LRG}}\leq 1.25$ and the stellar mass range of
$M_{\star,\mathrm{in}}\geq 10^{10.3}M_\odot$, where the stellar mass $M_{\star,\mathrm{in}}$ is derived by the stellar population synthesis fitting assuming the
\citet{1955ApJ...121..161S} IMF. As shown in
\citet{2026PASJ...78..416O}, their photometric redshifts are accurate and precise
with a scatter $\sigma_z\lesssim 0.02$ in the redshift range of
$0.4 \lesssim z_{\mathrm{LRG}} \lesssim 1$. In \citet{2026OJAp....961580F}, we select a photometric LRG subsample with $0.4 < z < 0.6$ and divide it into seven stellar mass bins spanning $10^{10.3}M_\odot < M_{\star,\mathrm{in}} < 10^{11.7}M_\odot$. In this paper, we focus on the highest-mass bin, $10^{11.5}M_\odot < M_{\star,\mathrm{in}} < 10^{11.7}M_\odot$, that contains $1.02\times10^4$ galaxies, and investigate it in more detail. 

\begin{table*}
  \tbl{List of 14 strong lens systems used for our analysis.\footnotemark[$*$] }{%
  \begin{tabular}{cccccc}
      \hline
Name & $z_{\mathrm{l}}$ & $z_{\mathrm{s}}$ & 
$\theta_\mathrm{E}$ [$\mathrm{arcsec}$] & $\sigma_{\theta_\mathrm{E}}$ [$\mathrm{arcsec}$] & Reference \\
      \hline
HSCJ022610$-$042011 & 0.495 & 1.232 & 1.036 & 0.104 & {HSC \citep{2019A&A...630A..71S}} \\
HSCJ022346$-$053418 & 0.499 & 1.444 & 1.469 & 0.147 & {HSC \citep{2019A&A...630A..71S}} \\
HSCJ144307$-$004056 & 0.5 & 1.07 & 1.096 & 0.110 & {HSC \citep{2019A&A...630A..71S}} \\
HSCJ015618$-$010747 & 0.542 & 1.167 & 0.841 & 0.084 & {HSC \citep{2019A&A...630A..71S}} \\
HSCJ141815+015832 & 0.556 & 2.139 & 1.395 & 0.140 & {HSC \citep{2019A&A...630A..71S}} \\
HSCJ120623+001507 & 0.563 & 3.12 & 1.165 & 0.117 & {HSC \citep{2019A&A...630A..71S}} \\
HSCJ141649+013822 & 0.434 & 1.313 & 1.13 & 0.22 & {HSC \citep{2025MNRAS.540.3384G}} \\
SL2SJ140546+524311 & 0.53 & 3.01 & 1.431 & 0.008 & {SL2S \citep{2024MNRAS.530.1474T}} \\
SDSSJ074724.12+505537.5 & 0.4384 & 0.8983 & 0.699 & 0.0025 & {BELLS \citep{2024MNRAS.530.1474T}} \\
SDSSJ134910.30+361239.7 & 0.4396 & 0.8926 & 0.673 & 0.0035 & {BELLS \citep{2024MNRAS.530.1474T}} \\
SDSSJ154503.57+274805.3 & 0.5218 & 1.2886 & 1.122 & 0.009 & {BELLS \citep{2024MNRAS.530.1474T}} \\
SDSSJ163150.33+185404.1 & 0.4081 & 1.0863 & 1.629 & 0.001 & {BELLS \citep{2024MNRAS.530.1474T}} \\
SDSSJ230335.17+003703.2 & 0.4582 & 0.9363 & 0.986 & 0.0035 & {BELLS \citep{2024MNRAS.530.1474T}} \\
      \hline
    \end{tabular}}\label{table:catalog_sl}
\begin{tabnote}
\footnotemark[$*$] For each strong lens system, we show the lens redshift $z_{\mathrm{l}}$, the source redshift $z_{\mathrm{s}}$, the Einstein radius $\theta_\mathrm{E}$, its $1\sigma$ uncertainty $\sigma_{\theta_\mathrm{E}}$, and the reference.
\end{tabnote}
\end{table*}

\subsection{Strong lensing data}\label{sec:strong_lens}
In this paper, we use galaxy-galaxy strong lens systems in the literature for which both lens and source galaxies have spectroscopic redshifts and the Einstein radii are measured. We further impose the requirement that redshifts of lens galaxies are within the same range as that of the weak lens sample, $0.4 < z < 0.6$, and that their stellar masses are also within the same range, $10^{11.5} M_\odot < M_{\star,\mathrm{in}} < 10^{11.7} M_\odot$. Our method of assigning stellar mass $M_{\star,\mathrm{in}}$ for each strong lens system is described in Section~\ref{sec:matching}. By applying these selection criteria, we obtain a sample of 13 strong lens systems, which are selected from the Survey of Gravitationally-lensed Objects in HSC Imaging \citep[SuGOHI;][]{2019A&A...630A..71S, 2025MNRAS.540.3384G}, the Strong Lensing Legacy Survey \citep[SL2S;][]{2012ApJ...761..170G,2024MNRAS.530.1474T}, and the BOSS Emission-Line Lens Survey \citep[BELLS;][]{2012ApJ...744...41B,2024MNRAS.530.1474T}. The details of our strong lens sample are summarized in Table~\ref{table:catalog_sl}.

For each strong lens system in the sample, we calculate the two-dimensional projected mass enclosed within the Einstein radius $M_\mathrm{2D}(<\theta_\mathrm{E})$, using the Einstein radius $\theta_\mathrm{E}$, the lens redshift $z_{\mathrm{l}}$, and the source redshift $z_{\mathrm{s}}$ as
\begin{equation}
  M_\mathrm{2D}(<\theta_\mathrm{E}) = \pi {D_{\mathrm{ol}}}^2(z_{\mathrm{l}}) {\theta_\mathrm{E}}^2\Sigma_\mathrm{cr}(z_{\mathrm{l}},\,z_{\mathrm{s}}),
  \label{eq:m2d}
\end{equation}
where $D_{\mathrm{ol}}(z_{\mathrm{l}})$ is the angular diameter distance between the observer and the lens galaxy at redshift $z_{\mathrm{l}}$ and $\Sigma_{cr}(z_{\mathrm{l}},\,z_{\mathrm{s}})$ is the critical surface mass density for lens and source redshifts of $z_{\mathrm{l}}$ and $z_{\mathrm{s}}$, respectively. By combining the $M_\mathrm{2D}(<\theta_\mathrm{E})$ measurements from different strong lens systems with different Einstein radii and redshifts, we obtain an average projected mass profile $M_\mathrm{2D}(<r)$ for the lensing galaxies. Since the spectroscopic redshifts $z_{\mathrm{l}}$ and $z_{\mathrm{s}}$ are determined with sufficient precision, we consider only the uncertainty in $\theta_\mathrm{E}$ when estimating the uncertainty in $M_\mathrm{2D}(<r)$, and propagate the uncertainty in $\theta_\mathrm{E}$ to $M_\mathrm{2D}(<r)$ following Equation~\eqref{eq:m2d}.

\subsection{Matching strong and weak lensing samples}\label{sec:matching}
As described in the previous section, we select strong lens systems with stellar masses in the same range as the weak lensing sample, $10^{11.5} M_\odot < M_{\star,\mathrm{in}} < 10^{11.7} M_\odot$. The specific procedure is as follows.
First, we cross-match the strong lens sample with the Dark Energy Spectroscopic Instrument (DESI) DR1 galaxy catalog \citep{2024ApJ...961..173Z}, which is expected to have substantial overlap with our sample given its wide footprint coverage, and obtain the stellar masses measured by DESI for each strong lens system. We then cross-match the HSC-SSP final-year photometric LRG sample \citep{2026PASJ...78..416O}, which is used in our weak lensing analysis, with the DESI DR1 galaxy catalog. We divide the DESI stellar masses into 20 bins over the range $10^{10}M_\odot < M_{\star\mathrm{,DESI}} < 10^{12}M_\odot$. For each DESI stellar mass bin, we count the number of objects falling within each bin of stellar masses measured by HSC. We then select the DESI stellar mass bin that contains the largest number of objects in the $10^{11.5}M_\odot < M_{\star\mathrm{,in}} < 10^{11.7}M_\odot$ HSC stellar mass bin and use the objects in this DESI bin for our analysis. This allows us to determine the DESI stellar mass range of $2\times10^{11} M_\odot \lesssim M_{\star,\mathrm{DESI}} < 10^{12} M_\odot$ corresponding to the HSC stellar mass range of $10^{11.5} M_\odot < M_{\star,\mathrm{in}} < 10^{11.7} M_\odot$.
This procedure takes account of the offset between the HSC stellar masses, which are derived from photometric data assuming the Salpeter IMF, and the DESI stellar masses, which are derived from spectroscopic data assuming the \citet{2003PASP..115..763C} IMF. We then select the strong lens systems whose DESI stellar masses are within the corresponding range to obtain the strong lens sample described in Section~\ref{sec:strong_lens}.

\subsection{Strong-lensing selection bias}\label{sec:bias}

\begin{figure}
 \begin{center}
   \includegraphics[width=8.5cm]{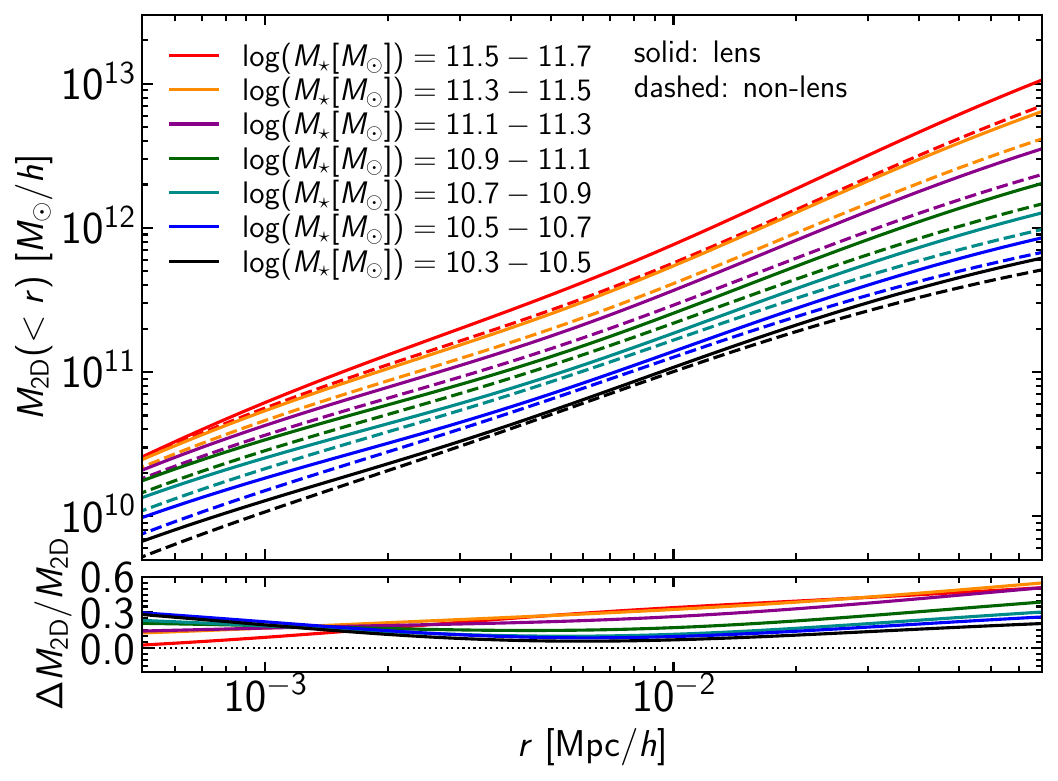} \\
   \includegraphics[width=8.5cm]{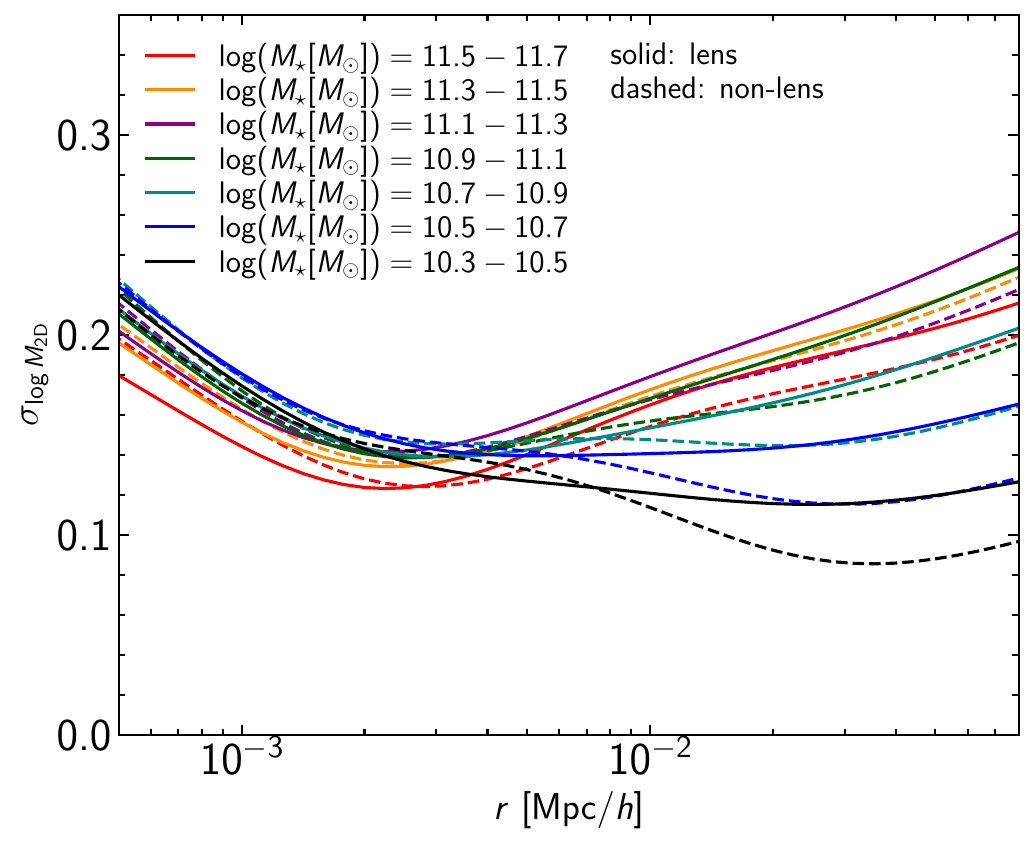} 
 \end{center}
\caption{{\it Upper:} The average enclosed projected masses $M_{\mathrm{2D}}(<r)$ for lens ({\it solid}) and non-lens ({\it dashed}) samples derived from mock catalogs of \citet{2025OJAp....8E...8A}. We show results for 7 different stellar mass bins, with enclosed projected masses being higher for higher stellar mass bins. The sub-panel at the bottom shows bias values $\Delta M_{\mathrm{2D}}/M_{\mathrm{2D}}=\{M_{\mathrm{2D}}(\mbox{lens})-M_{\mathrm{2D}}(\mbox{non-lens})\}/M_{\mathrm{2D}}(\mbox{non-lens})$ as a function of radius $r$. {\it Lower:} Similar to the upper panel, but intrinsic scatters of enclosed projected masses $\sigma_{\log M_{\mathrm{2D}}}$ are plotted.
{Alt text: The x axis shows the radius from 0.0005 to 0.08 megaparsec per h. The y axis shows the enclosed mass and the intrinsic scatter.}
}\label{fig:mcyl} 
\end{figure}

Lensing galaxies of strong lens systems sometimes represent a highly biased population of galaxies because of the strong dependence of the lensing cross section on the internal structure of galaxies. For instance, assuming a singular isothermal sphere, the lensing cross section has a steep dependence on the velocity dispersion $\sigma$ as $\sigma^4$, and therefore strong lensing preferentially selects galaxies with high velocity dispersions for a given galaxy sample. Indeed previous studies show that such a strong-lensing selection effect is significant \citep[e.g.,][]{2009MNRAS.398..635M,2016MNRAS.462.3255C,2023A&A...678A...4S}.

Here we estimate and correct for the strong-lensing selection bias using a mock catalog of \citet{2025OJAp....8E...8A}. This mock catalog is constructed based on the halo model approach, for which mass models of individual galaxies are modeled by the suppression of a dark matter halo component modeled by the \citet[][hereafter NFW]{1997ApJ...490..493N} profile and a stellar mass component modeled by the \citet{1990ApJ...356..359H} profile. We use the mock catalog of lensed quasars in the Legacy Survey of Space and Time (LSST) with the Rubin Observatory, which is constructed assuming the Salpeter IMF, for estimating the strong-lensing selection bias. While in principle the source population and the survey depth should be matched to those of the strong lens sample used in the analysis, we adopt this mock catalog as a proxy because our strong lens sample is heterogeneous and is selected from several different surveys, and as a result deriving the exact selection bias for such sample is technically challenging. We expect that our current approach still enables us to estimate a reasonable first-order approximation of the strong-lensing selection bias.

In order to derive the strong-lensing selection bias, we extract lens model parameters, such as halo masses, concentration parameters, stellar masses, and galaxy sizes, from the mock catalog. In addition, we generate a mock realization of these model parameters for a non-lens sample without the strong-lens selection for a small sky area of 100~deg$^2$. For both lens and non-lens galaxy samples,  we first add measurement noises to stellar masses of individual galaxies assuming the log-normal distribution with the scatter $\sigma_{\ln M_\star}=0.31$, which corresponds to a typical measurement error of stellar masses of low-redshift galaxies from optical broadband photometry \citep[e.g.,][]{2014ApJS..210....3M}, and use these ``observed'' stellar masses for defining subsample. On the other hand, we ignore the effect of the photometric redshift uncertainty given its small scatter \citep{2026PASJ...78..416O}. We then apply the sample stellar mass and redshift cuts of $10^{11.5}M_\odot<M_{\star}<10^{11.7}M_\odot$ and $0.4<z<0.6$ to construct subsamples. For each subsample, we compute an average of the enclosed projected mass $\log M_{\mathrm{2D}}(<r)$, which is a quantity that can be directly constrained from observations of the Einstein radii (see Section~\ref{sec:strong_lens}), as a function of $r$, and derive the bias by the ratio of these two averages of the enclosed projected masses. Figure~\ref{fig:mcyl} shows enclosed projected masses of the lens and non-lens samples as a function of the radius. We find that, in the strong lensing region ($r\sim 10^{-3}-10^{-2}\mathrm{Mpc}/h$), $M_{\mathrm{2D}}$ is biased high at $\sim 30\%$. Our result also exhibits the radius dependence of the bias such that the bias is higher at larger radii. This result suggests that strong-lensing selection bias should be properly taken into account when combining stacked strong and weak lensing results. In Figure~\ref{fig:mcyl}, we also show results for different stellar mass bins, which reveal a complex dependence of the strong-lensing selection bias on stellar masses. 

In addition to the strong-lensing selection bias, we derive an intrinsic scatter of the enclosed projected mass using the same mock catalog analysis. For a given stellar mass and redshift of a galaxy, the halo model of \citet{2025OJAp....8E...8A} predicts such intrinsic scatter of the enclosed projected mass originating from the scatter of the stellar mass--halo mass relation, the scatter of the stellar mass--galaxy size relation, and the scatter of the concentration parameter. Figure~\ref{fig:mcyl} indicates that the intrinsic scatter is typically $\sigma_{\log M_{\mathrm{2D}}}\sim 0.14$ in the strong lensing region. It is found that the intrinsic scatters for the lens and non-lens samples are similar, and that the stellar mass dependence is not very large, at least when restricted to the strong lensing region.

{As described above, our method for estimating the strong-lensing selection bias and the intrinsic scatter is an approximate one with simplified assumptions. For assessing the associated systematic uncertainty, we repeat our analysis with the stellar mass measurement scatter set to $\sigma_{\ln M_\star} = 0.55$, which corresponds to the stellar mass uncertainty reported by \citet{2014ApJS..210....3M} when accounting for the influence of SED modeling assumptions. The corresponding strong-lensing selection bias is $\Delta M_\mathrm{2D} / M_\mathrm{2D} \sim 0.6$ and the corresponding intrinsic scatter in the enclosed projected mass is $\sigma_{\log M_\mathrm{2D}} \sim 0.2$. The results of this analysis are presented in Table~\ref{table:para_coredpl}. The best-fit parameter values are consistent with those obtained using our fiducial scatter within $1\sigma$, indicating that our conclusions are not significantly affected. Nevertheless, we caution that our results carry systematic uncertainties at this level.}

\subsection{Fitting method}\label{sec:fitting}
We simultaneously fit the weak and strong lensing data. For fitting the weak lensing data, we follow \citet{2026OJAp....961580F} to perform a chi-square analysis using
\begin{align}
  \chi^2_\mathrm{WL} = \sum_{i=1}^{N} \frac{\left[\Delta\Sigma_{\mathrm{obs},i} - f_{\mathrm{model}}(r_i)\right]^2}{\sigma_{i}^2},& \label{eq:chi2_wl}\\
  f_{\mathrm{model}}(r)=\frac{\Delta\Sigma(r)}{1-\langle\Sigma_{\mathrm{cr}}^{-1}\rangle\Sigma(r)}&,
\end{align}
where $\Delta\Sigma_{\mathrm{obs},i}$ and $\sigma_i$ are the differential surface mass density and its uncertainty in the $i$-th radial bin obtained from the weak lensing measurement, respectively. {We ignore off-diagonal terms of the covariance matrix, which is justified by the fact that off-diagonal terms are negligibly small within the range of the radial bins of our interest where the shape noise is dominated \citep[e.g.,][]{2017MNRAS.471.3827S,2019PASJ...71..107M,2019ApJ...875...63M}.} The model-predicted surface mass density, surface mass density, and the average inverse critical surface density are denoted by $\Delta\Sigma(r)$, $\Sigma(r)$, and $\langle\Sigma_{\mathrm{cr}}^{-1}\rangle$, respectively. The models used in the fitting are described in the following section. The fitting is performed only for the inner part of the weak lensing data defined by
the range $0.013\,\mathrm{Mpc}/h<r <0.1\,\mathrm{Mpc}/h$, where the inner boundary is determined based on various systematic tests.  For more details of the fitting procedure for the weak-lensing data, see \citet{2026OJAp....961580F}.

For the strong lensing data, we need to account for the fact that the uncertainties in $M_\mathrm{2D}(<r)$ and $r$ are correlated because both are derived from the uncertainty in $\theta_\mathrm{E}$ as described in Section~\ref{sec:strong_lens}. We therefore perform a chi-square analysis using the distance between the model line and each observed data point along the direction of its error bar. Let the observed radius and projected mass be $r_\mathrm{obs}$ and $M_\mathrm{2D,obs}(<r_\mathrm{obs})$, respectively. In the log-log plane with $r$ on the $x$-axis and $M_\mathrm{2D}(<r)$ on the $y$-axis, the equation of the line passing through each data point and its error bar is given by
\begin{equation}
y=2(x-\log{r_\mathrm{obs}})+\log{M_\mathrm{2D,obs}(<r_\mathrm{obs})}.
\end{equation}
On the other hand, assuming an isothermal mass profile, the model prediction can be approximated as $M_\mathrm{2D}(<r)\propto r$, corresponding to
\begin{equation}
y=x-\log{r_\mathrm{obs}}+\log{M_\mathrm{2D}(<r_\mathrm{obs})}.
\end{equation}
We therefore define the strong lensing chi-square using the distance between the observed data point and the intersection of these two relations as
\begin{equation}
\chi^2_\mathrm{SL} = \sum_j \frac{5\left[\log{M_\mathrm{2D}(<r_{\mathrm{obs},j})}-\log{M_\mathrm{2D,obs}(<r_{\mathrm{obs},j})}\right]^2}{{\sigma_j}^2+{\sigma_\mathrm{int}}^2},
\end{equation}
where $j$ labels the strong lensing systems. $\sigma_j$ and $\sigma_\mathrm{int}$ represent the measurement uncertainty of each data point and the intrinsic scatter, respectively, along the direction of the error bar. The measurement error is $\sigma_j=\sqrt{\sigma_{\log r_{\mathrm{obs}}}^2+\sigma_{\log M_{\mathrm{2D,obs}}}^2}$.
The intrinsic scatter $\sigma_\mathrm{int}$ is expressed using the intrinsic scatter of $\log M_{\mathrm{2D}}$ inferred from the mock catalog (see Section~\ref{sec:bias}), $\sigma_{\log{M_\mathrm{2D}}}=0.14$,  as $\sigma_\mathrm{int}=\sqrt5\,\sigma_{\log{M_\mathrm{2D}}}$. {We note that it is also possible to directly predict the Einstein radius from the model and define the chi-square using the difference between the predicted and observed Einstein radii. However, this approach would complicate the incorporation of the intrinsic scatter and the strong-lensing selection bias correction into the likelihood. This is because redefining the intrinsic scatter and the strong-lensing selection bias as quantities associated with the Einstein radius would require applying individual corrections to each of the 13 lens systems, given that the mapping between $M_\mathrm{2D}$ and $\theta_\mathrm{E}$ depends on $\Sigma_\mathrm{cr}$ and hence on $z_\mathrm{l}$ and $z_\mathrm{s}$ of each system.}

Finally, we perform the joint fit by minimizing the total chi-square
\begin{equation}
\chi^2 = \chi^2_\mathrm{WL}+\chi^2_\mathrm{SL},
\end{equation}
and derive constraints on model parameters.

\section{Stellar and dark matter distributions}\label{sec:smdm}

\subsection{Mass models}

We fit the strong and weak lensing data to the two-component mass model of \citet{2026OJAp....961580F}, which consists of stellar and dark matter components. The stellar matter component is assumed to follow the Hernquist density profile \citep{1990ApJ...356..359H}
\begin{equation}
  \rho(r) = \frac{M_\star}{2\pi}\frac{a}{r}\frac{1}{\left(r+a\right)^3},
  \label{eq:rhosm}
\end{equation}
where $M_\star$ is the stellar mass and $a$ is the scale radius that is related to the effective (half-light) radius $r_\mathrm{e}$ as
\begin{equation}
    r_\mathrm{e} \approx 1.8153a.
\end{equation}
Analytic expressions of lensing profiles of the Hernquist density profile are provided in \citet{2001astro.ph..2341K}.

The dark matter component is assumed to follow a power-law profile with a central core. Specifically, we define the three-dimensional density profile as
\begin{equation}
  \rho_{\mathrm{DM}}\left(r\right)
  = M_\star\frac{3+\gamma}{2\pi^{3/2} \, r_\mathrm{e}^{3}} \frac{\Gamma\!\left(-\tfrac{\gamma}{2}\right)}{\Gamma\!\left(-\tfrac{1+\gamma}{2}\right)}A\left( \frac{r^{2} + r_\mathrm{c}^{2}}{r_\mathrm{e}^{2}} \right)^{\gamma/2},
\end{equation}
where $\Gamma(x)$ is the gamma function, $\gamma$ is the radial slope, $r_\mathrm{c}$ is the core radius, and $A$ is the dimensionless normalization parameter. Analytic expressions of lensing profiles of this cored power-law model are provided in \citet{2026OJAp....961580F}.

To explore the dependence of our results on the choice of the dark matter density profile, in addition to the cored power-law profile used in \citet{2026OJAp....961580F}, we also consider a broken power-law model in \citet{2024MNRAS.533..795K} that is more flexible. This broken power-law model has the projected mass density profile
\begin{equation}
\Sigma_\mathrm{DM}(r)=
\left\{
\begin{aligned}
  &\Sigma_\mathrm{B}(r/r_\mathrm{B})^{\gamma_\mathrm{in}} \;\;\;\;\;\mathrm{for}\;\; r\leq r_\mathrm{B},\\
  &\Sigma_\mathrm{B}(r/r_\mathrm{B})^{\gamma_\mathrm{out}} \;\;\;\;\;\mathrm{for}\;\; r\geq r_\mathrm{B},
\end{aligned}
\right.
\end{equation}
where $r_\mathrm{B}$ is the break radius, $\Sigma_\mathrm{B}$ is the surface mass density at the break radius, $\gamma_\mathrm{in}$ and $\gamma_\mathrm{out}$ are the inner and outer two-dimensional logarithmic slopes, respectively. The two-dimensional enclosed mass can be expressed in terms of the surface mass density as
\begin{align}
  &M_{2\mathrm{D},\mathrm{DM}}\left(<r\right) =\int_{0}^{r} 2\pi r'\Sigma_{\mathrm{DM}}\left(r'\right)dr' \nonumber\\
  =& 
  \left\{
  \begin{aligned}
 &\frac{2\pi\Sigma_\mathrm{B}r_\mathrm{B}^2}{\gamma_\mathrm{in}+2}\left(\frac{r}{r_\mathrm{B}}\right)^{\gamma_\mathrm{in}+2} \;\;\;\;\;\mathrm{for}\;\; r\leq r_\mathrm{B},\\
  &2\pi\Sigma_\mathrm{B}r_\mathrm{B}^2\left[\frac{1}{\gamma_\mathrm{in}+2}+\frac{(r/r_\mathrm{B})^{\gamma_\mathrm{out}+2}-1}{\gamma_\mathrm{out}+2}\right] \;\;\;\;\;\mathrm{for}\;\; r\geq r_\mathrm{B}.
\end{aligned}
\right.
\end{align}
In addition, the average surface mass density within the radius $r$ is given by
\begin{align}
  &\bar{\Sigma}_{\mathrm{DM}} \left(<r\right)=\frac{M_{\mathrm{DM},2\mathrm{D}}\left(<r\right)}{\pi r^{2}}\nonumber\\
  =& 
  \left\{
  \begin{aligned}
 &\frac{2\Sigma_\mathrm{B}}{\gamma_\mathrm{in}+2}\left(\frac{r}{r_\mathrm{B}}\right)^{\gamma_\mathrm{in}}\;\;\;\;\;\mathrm{for}\;\; r\leq r_\mathrm{B},\\
  &\frac{2\Sigma_\mathrm{B}r_\mathrm{B}^2}{r^2}\left[\frac{1}{\gamma_\mathrm{in}+2}+\frac{(r/r_\mathrm{B})^{\gamma_\mathrm{out}+2}-1}{\gamma_\mathrm{out}+2}\right]\;\;\;\;\;\mathrm{for}\;\;r\geq r_\mathrm{B}.
\end{aligned}
\right.
\end{align}
From these results, the differential surface mass density is computed as
\begin{align}
  &\Delta \Sigma_{\mathrm{DM}}\left(r\right) =\bar{\Sigma}_{\mathrm{DM}}\left(<r\right) - \Sigma_{\mathrm{DM}}\left(r\right)\nonumber\\
  =& 
  \left\{
  \begin{aligned}
 &-\frac{\gamma_\mathrm{in}\Sigma_\mathrm{B}}{\gamma_\mathrm{in}+2}\left(\frac{r}{r_\mathrm{B}}\right)^{\gamma_\mathrm{in}}\;\;\;\;\;\mathrm{for}\;\;r\leq r_\mathrm{B},\\
  &\frac{2\Sigma_\mathrm{B}r_\mathrm{B}^2}{r^2}\left[\frac{1}{\gamma_\mathrm{in}+2}+\frac{(r/r_\mathrm{B})^{\gamma_\mathrm{out}+2}-1}{\gamma_\mathrm{out}+2}\right]
  -\Sigma_\mathrm{B}\left(\frac{r}{r_\mathrm{B}}\right)^{\gamma_\mathrm{out}}\\ & \;\;\;\;\;\mathrm{for}\;\; r\geq r_\mathrm{B}.
\end{aligned}
\right.
\end{align}

\begin{figure*}
 \begin{center}
   \includegraphics[width=5.8cm]{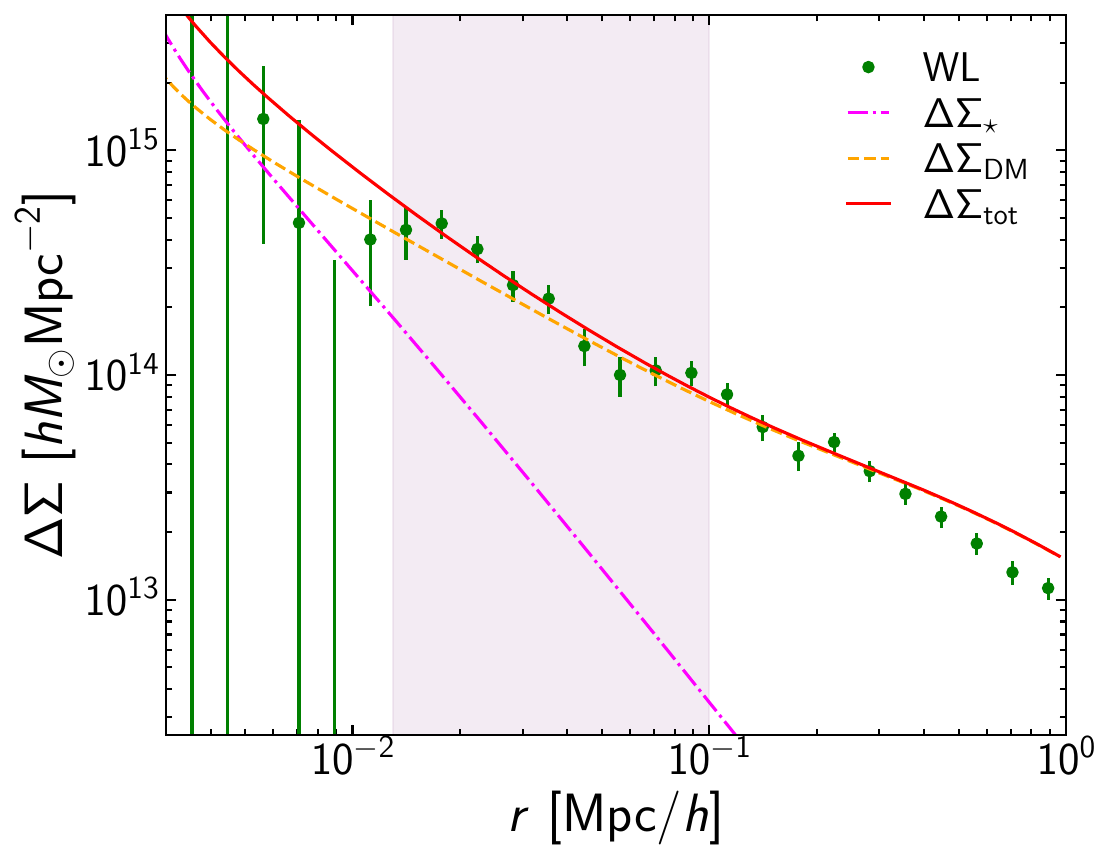} 
   \includegraphics[width=5.8cm]{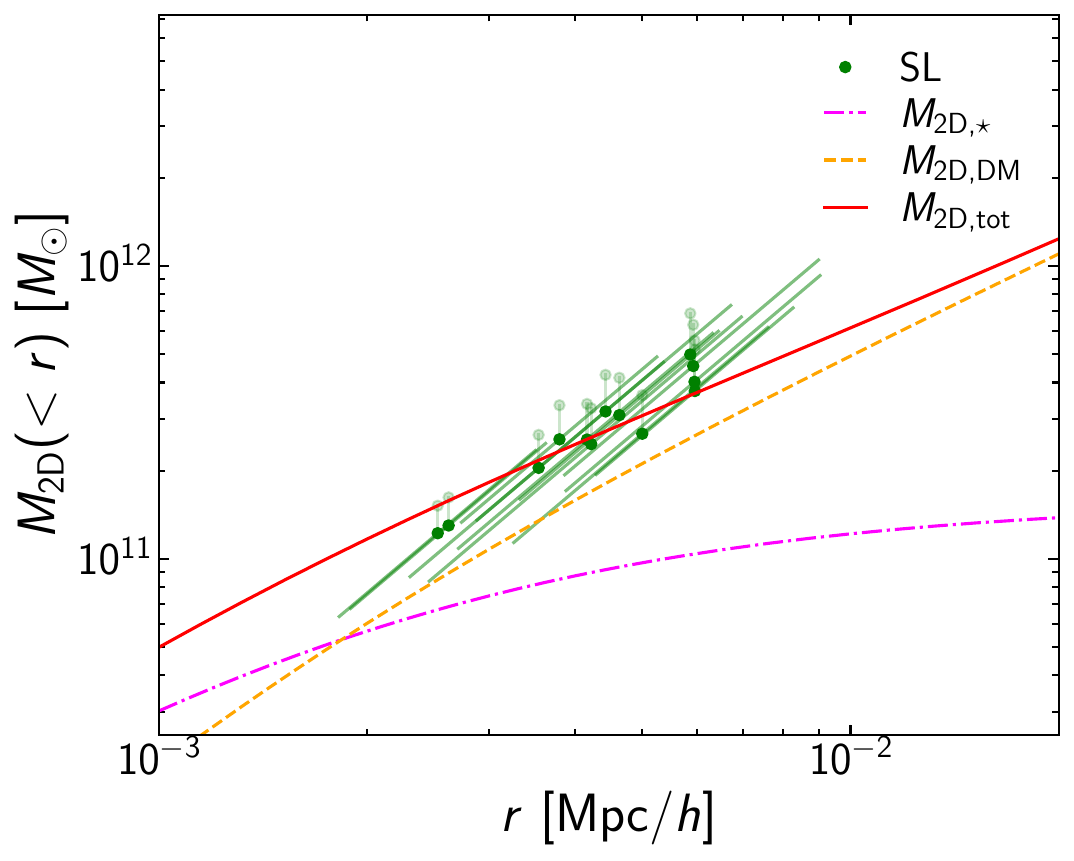} 
   \includegraphics[width=5.8cm]{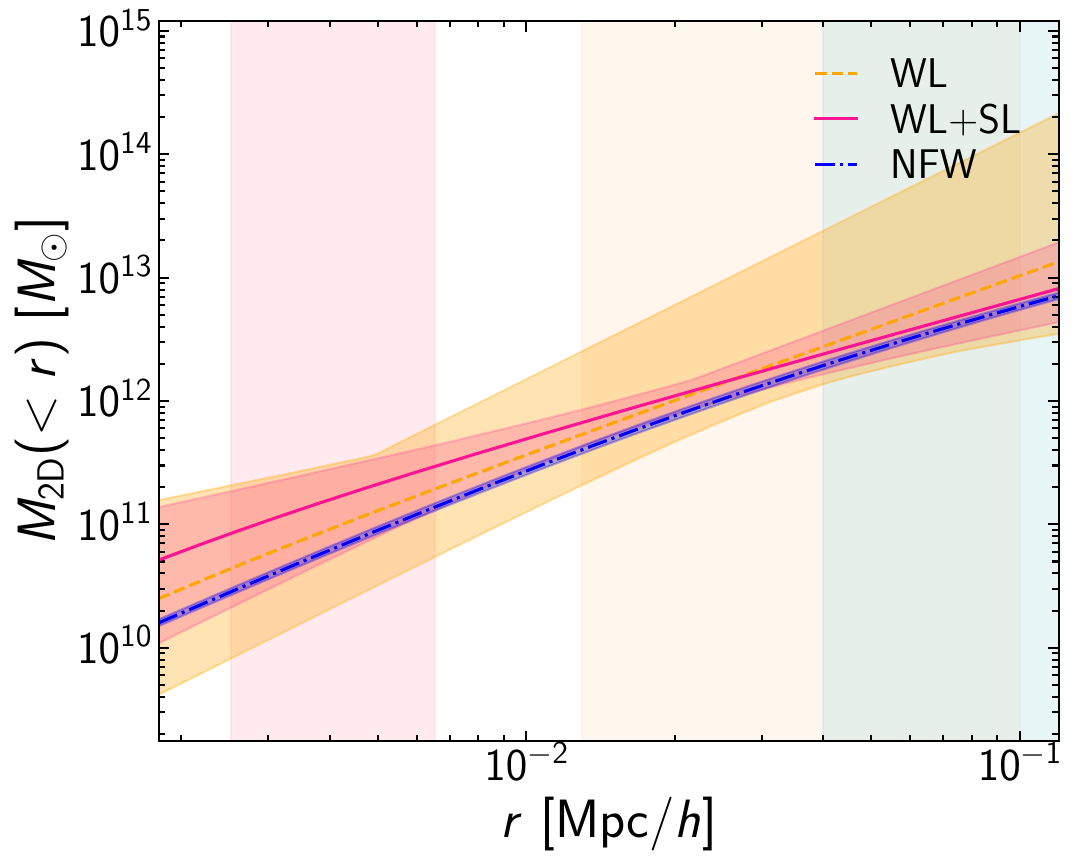}
   \end{center}
   \begin{flushleft}
   \includegraphics[width=5.8cm]{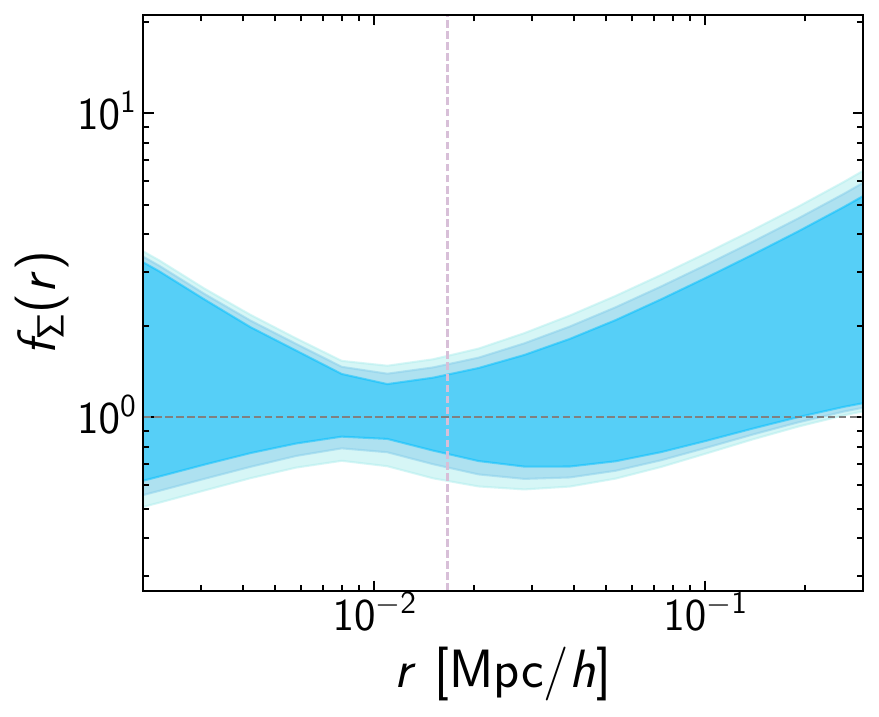} 
   \includegraphics[width=5.8cm]{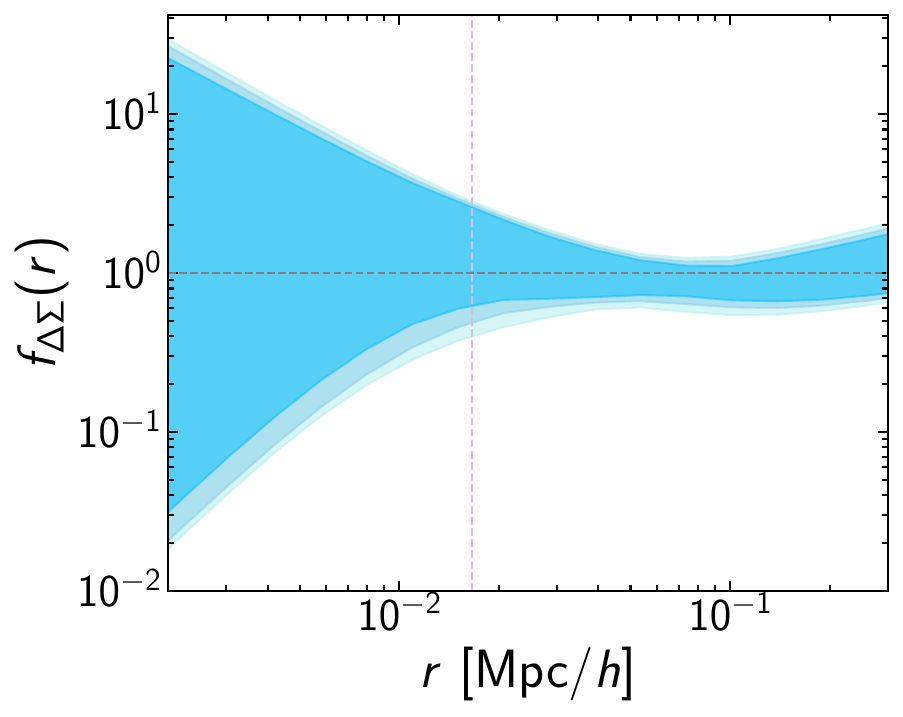} 
\end{flushleft}
 \caption{{\it Upper left:} Differential surface mass density profile obtained from the weak lensing measurement and the result from joint fit to the weak and strong lensing data using the cored power-law model for the dark matter component. The green points with errors show the stacked weak lensing measurement \citep{2026OJAp....961580F}.  The magenta dash-dotted, the orange dashed, and the red solid lines show best-fit profiles of the stellar matter component, the dark matter component, and the total mass, respectively. The shaded region denotes the fitting range of the weak lensing analysis. {\it Upper middle:} Enclosed projected mass profile obtained from the strong lensing measurements and the result from joint fit to the weak and strong lensing data, using the cored power-law model for the dark matter component. The green points with errors show the observed enclosed projected masses measured from the strong lens sample, after correcting for the strong-lensing selection bias (see Section~\ref{sec:bias}). Their error bars include both statistical uncertainties in the Einstein radii and the intrinsic scatter. The light-colored points show observed values before correcting for the strong-lensing selection bias. {\it Upper right:} Constraints on the enclosed projected mass profile. The orange dashed and {red} solid lines show the results from the weak lensing analysis and the joint fit to the weak and strong lensing measurements, respectively. The blue dash-dotted line indicates the NFW profile obtained from the outer profile fitting. The shaded regions show the $1\sigma$ uncertainties. The orange, {red}, and blue shaded radial regions indicate the fitting ranges of the weak lensing analysis, the strong lensing analysis, and the outer NFW fit, respectively. {\it Lower left:} The ratio of the surface mass densities $f_{\Sigma}(r)$ defined in Equation~\eqref{eq:fsigma} constrained from {the joint fit to the weak and strong lensing measurements.} The shaded regions represent the $1\sigma$, $2\sigma$, and $3\sigma$ confidence intervals from darker to lighter colors. The vertical dashed line shows $r=5r_\mathrm{e}$. {\it Lower middle:} Same as the lower left panel, but for $f_{\Delta\Sigma}(r)$ defined in Equation~\eqref{eq:fdsigma}.
 {Alt text: The x axis shows the radius. The y axis shows the differential surface density, the enclosed mass and the ratios.}
 }\label{fig:coredpl} 
\end{figure*}

\subsection{Results}

Figure~\ref{fig:coredpl} compares the best-fitting model from the joint strong and weak lensing analysis using the cored power-law model for the dark matter component with the weak and strong lensing data. The best-fit parameters are summarized in Table~\ref{table:para_coredpl}. We find that the total mass profile provides a good fit to the observed data points. Figure~\ref{fig:coredpl} also shows constraints on the enclosed projected mass profile with $1\sigma$ uncertainties. By combining the weak and strong lensing measurements, we obtain the tighter constraint than that obtained from the weak lensing analysis alone.

\begin{table*}
  \tbl{Best-fit parameters for weak lensing only and for the joint weak and strong lensing analysis using the cored power-law model.\footnotemark[$*$] }{%
  \begin{tabular}{ccccccccc}
      \hline
Data & {$\sigma_{\ln{M_{\star,\mathrm{DESI}}}}$} & {$\sigma_{\log{M_\mathrm{2D}}}$} & $r_{\mathrm{e}}$ [$\mathrm{Mpc}/h$]& 
$A$ & $\gamma$ & $r_{\mathrm{c}}$ [$\mathrm{Mpc}/h$] & $M_{\star,\mathrm{fit}}$ [$M_\odot / h$]
& $\chi^2/\mathrm{dof}$ \\
            \hline
WL & & & $3.3 \times 10^{-3}$ & $0.2^{+8.0}_{-0.1}$ & $-1.56^{+0.55}_{-0.52}$ & $1^{+16}_{-1} \times 10^{-3}$ & $2.7^{+1.6}_{-2.5} \times 10^{11}$ & 6.4/5  \\
WL+SL & {$0.31$} & {$0.14$} & $3.3 \times 10^{-3}$ & $1.0^{+29.0}_{-0.6}$ & $-1.90^{+0.25}_{-0.26}$ & $1^{+4}_{-1} \times 10^{-3}$ & $1.1^{+1.2}_{-1.0} \times 10^{11}$ & 9.9/18 \\
{WL+SL} & {$0.55$} & {$0.20$} & {$3.3 \times 10^{-3}$} & {$33^{+7}_{-25}$} & {$-2.40^{+0.30}_{-0.10}$} & {$6^{+2}_{-2} \times 10^{-3}$} & {$1.5^{+3.5}_{-0.9} \times 10^{10}$} & {13.7/18} \\
      \hline
    \end{tabular}}\label{table:para_coredpl}
\begin{tabnote}
\footnotemark[$*$]  The errors are shown at the $1\sigma$ level. {The case $\sigma_{\ln{M_{\star,\mathrm{DESI}}}}=0.31$ corresponds to our fiducial choice, and and $\sigma_{\ln{M_{\star,\mathrm{DESI}}}}=0.55$ is for a systematics test.}
\end{tabnote}
\end{table*}

\begin{table*}
  \tbl{Best-fit parameters for the joint weak and strong lensing analysis using the broken power-law model.\footnotemark[$*$] }{%
  \begin{tabular}{ccccccc}
      \hline 
      Data &
$r_{\mathrm{Ein}}$ [$\mathrm{Mpc}/h$] & $\gamma_\mathrm{in}$ & $\gamma_\mathrm{out}$ & $r_{\mathrm{B}}$ [$\mathrm{Mpc}/h$] & $M_{\star,\mathrm{fit}}$ [$M_\odot / h$]
& $\chi^2/\mathrm{dof}$ \\
            \hline
WL+SL & $2.5^{+2.0}_{-1.0} \times 10^{-3}$ & $-0.03^{+0.03}_{-0.63}$ & $-0.80^{+0.00}_{-0.19}$ & $5^{+5}_{-3} \times 10^{-3}$ & $2.6^{+0.6}_{-0.4} \times 10^{11}$ & 13.8/17 \\
      \hline
    \end{tabular}}\label{table:para_brokenpl}
\begin{tabnote}
\footnotemark[$*$] The errors are shown at the $1\sigma$ level.
\end{tabnote}
\end{table*}

\begin{figure*}
 \begin{center}
   \includegraphics[width=5.8cm]{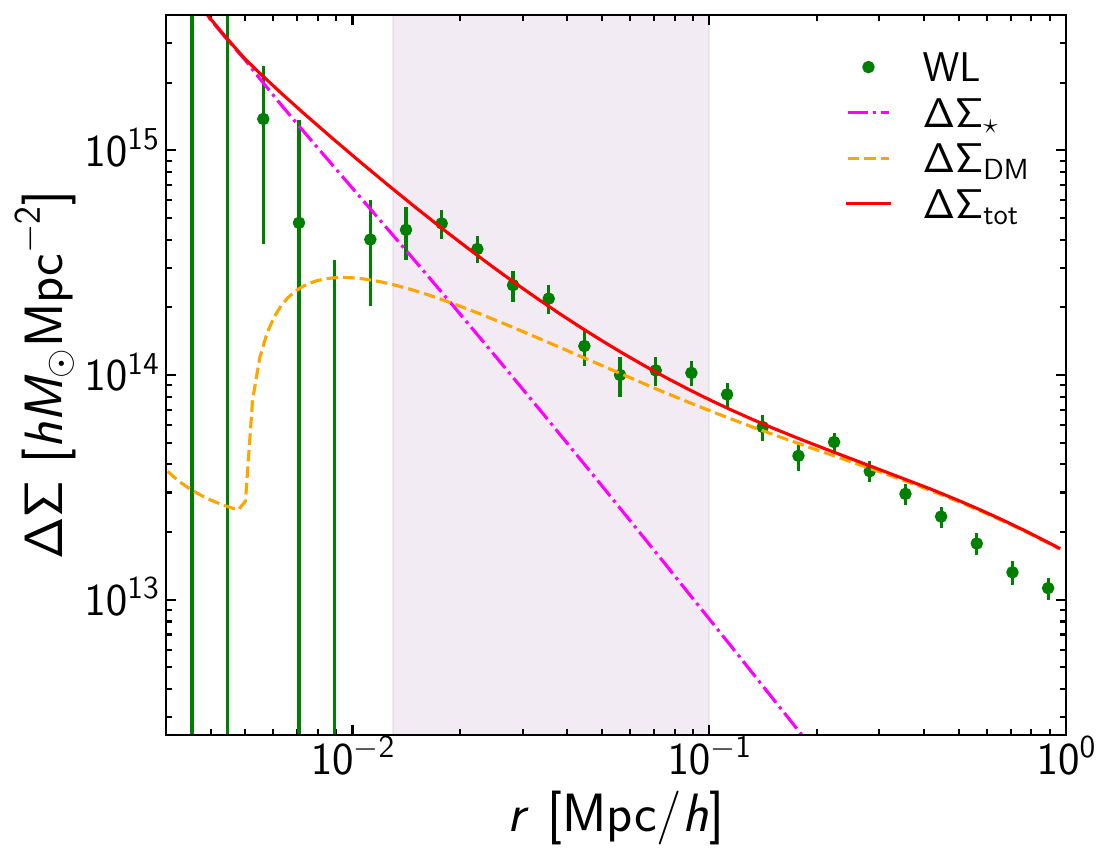} 
   \includegraphics[width=5.8cm]{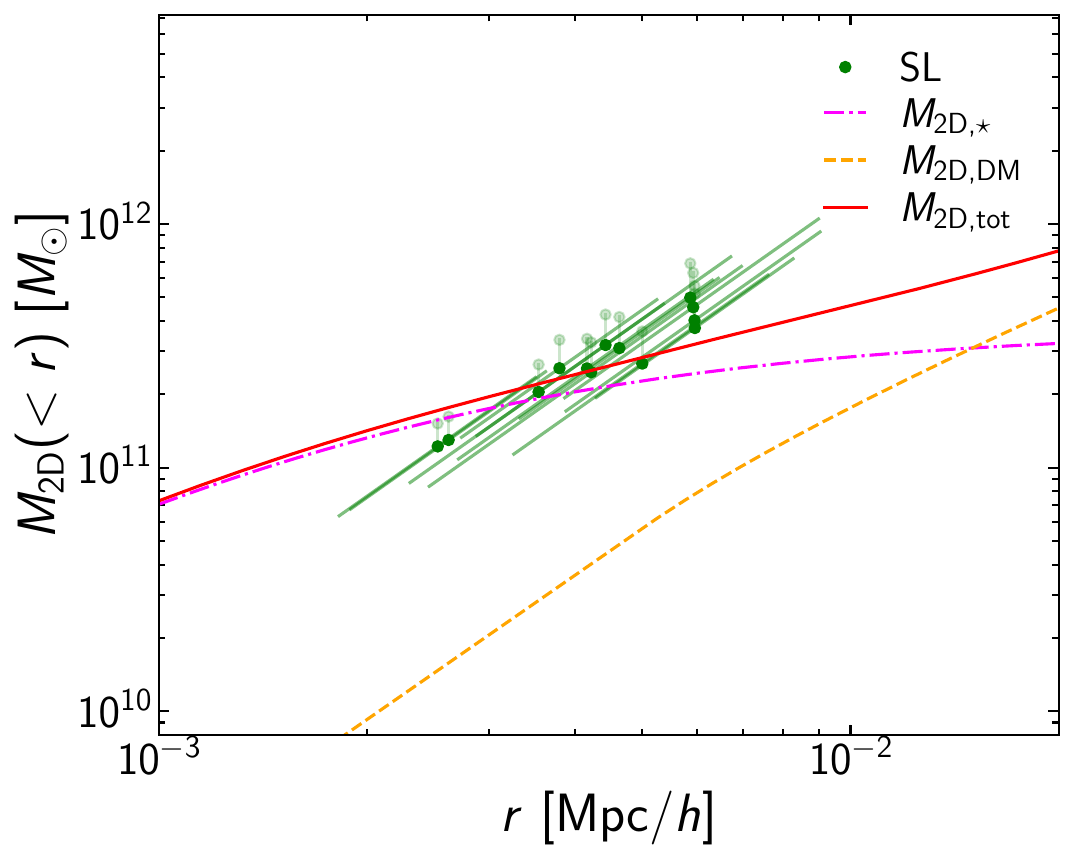} 
   \includegraphics[width=5.8cm]{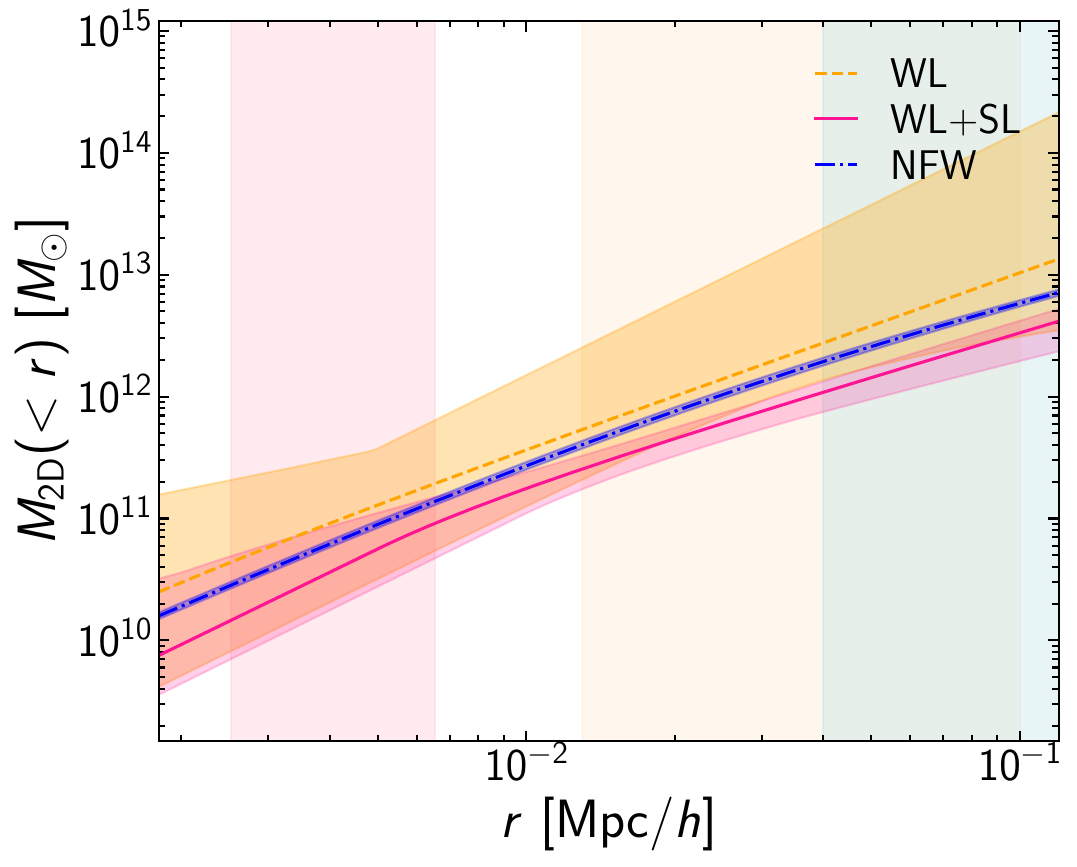}
   \end{center}
   \begin{flushleft}
   \includegraphics[width=5.8cm]{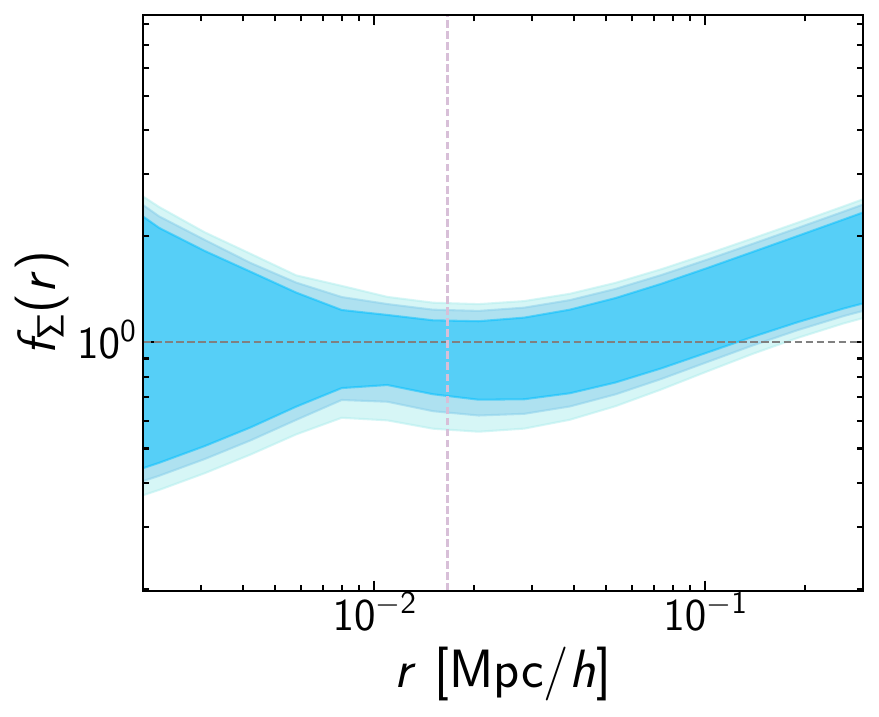} 
   \includegraphics[width=5.8cm]{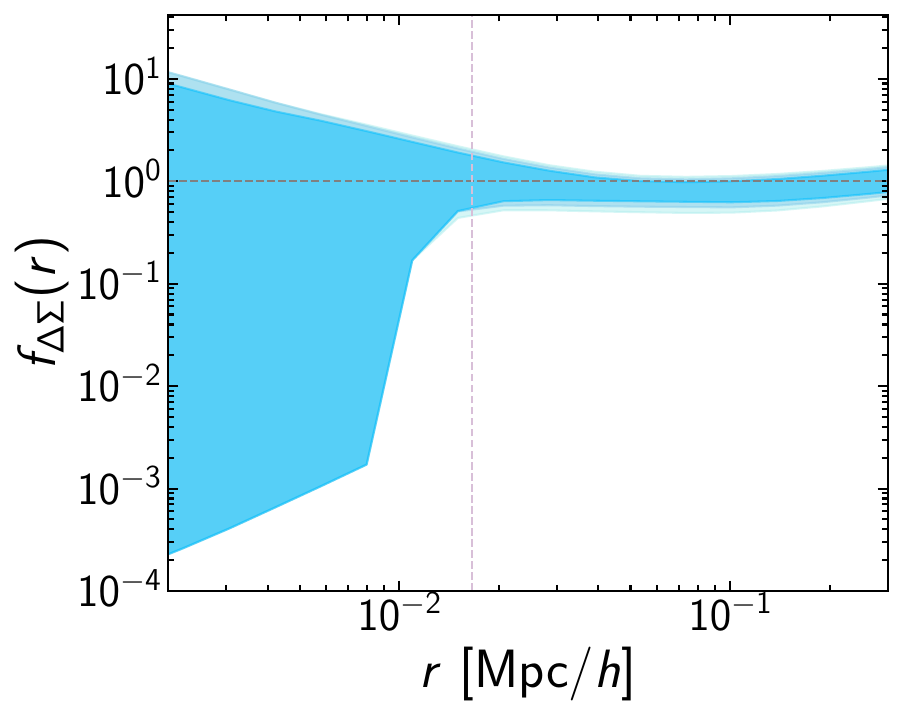} 
\end{flushleft}
 \caption{Same as Figure~\ref{fig:coredpl}, but for the broken power-law model for the dark matter distribution. {Alt text: The x axis shows the radius. The y axis shows the differential surface density, the enclosed mass and the ratios.}}\label{fig:brokenpl} 
\end{figure*}

As in \citet{2026OJAp....961580F}, it is useful to compare our constraints on the central dark matter density profile with the NFW profile, where the halo mass of the NFW profile is determined by fitting the outer part of the profile obtained from the weak lensing analysis over the range $0.04\,\mathrm{Mpc}/h < r < 3\,\mathrm{Mpc}/h$ \citep[see][for details]{2026OJAp....961580F}. We quantify the difference between the two profiles using their ratios. Specifically, we define
\begin{equation}
f_{\Sigma}(r) = \frac{\Sigma_{\mathrm{DM}}(r)}{\Sigma_{\mathrm{NFW}}(r)},
\label{eq:fsigma}
\end{equation}
as the ratio of the surface mass densities, and
\begin{equation}
f_{\Delta\Sigma}(r) = \frac{\Delta\Sigma_{\mathrm{DM}}(r)}{\Delta\Sigma_{\mathrm{NFW}}(r)},
\label{eq:fdsigma}
\end{equation}
as the ratio of the differential surface mass densities. We show $f_{\Sigma}(r)$ and $f_{\Delta\Sigma}(r)$ in Figure~\ref{fig:coredpl}. Both $f_{\Sigma}(r)$ and $f_{\Delta\Sigma}(r)$ are consistent with unity within the $1\sigma$ uncertainties, indicating that the central dark matter distribution obtained in this work is consistent with the NFW profile. This result is consistent with previous strong lensing analyses \citep[e.g.,][]{2014MNRAS.439.2494O,2015ApJ...800...94S,2015ApJ...814...26N,2021MNRAS.503.2380S,2025MNRAS.541....1S,2026MNRAS.548ag683L}, and is also consistent with that obtained from the weak lensing analysis alone \citep{2026OJAp....961580F}.

Using the broken power-law model, we obtain results consistent with those obtained using the cored power-law model. The best-fit parameters are listed in Table~\ref{table:para_brokenpl}. The fitting results for the weak and strong lensing data as well as the constraints on the enclosed projected mass profile
are shown in Figure~\ref{fig:brokenpl}. As in the cored power-law case, we compare the resulting dark matter distribution with the NFW profile obtained by fitting the outer part of the density profile and show the results also in Figure~\ref{fig:brokenpl}.

The consistency of the results obtained using different dark matter distribution models demonstrates the robustness of our results. Furthermore, despite the greater flexibility of the broken power-law model compared with the cored power-law model, we do not find an improvement in the goodness of fit. This supports the conclusion that a single power-law model provides a sufficiently good description of the data over the limited radial range probed in this study.

\begin{figure}
 \begin{center}
   \includegraphics[width=8.6cm]{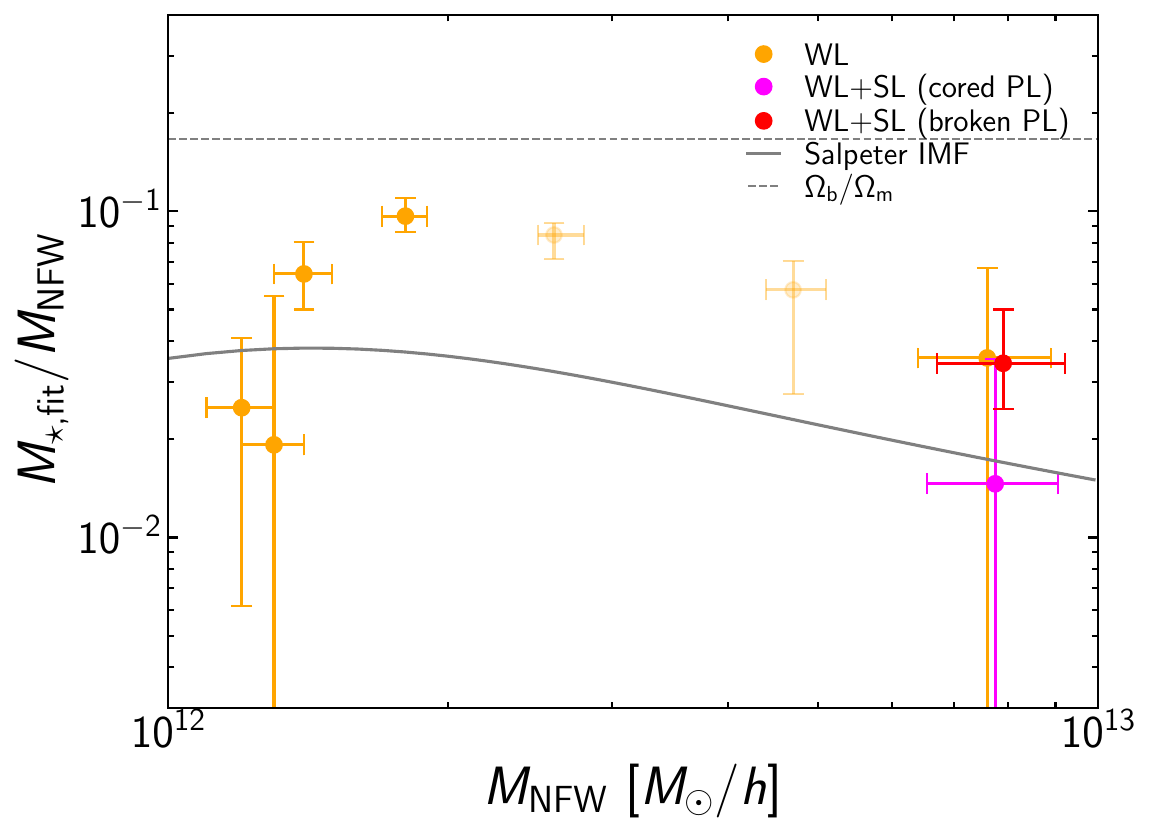} 
 \end{center}
\caption{The constraint on the SHMR. The vertical and horizontal axes indicate the stellar mass $M_{\star,\mathrm{fit}}$ obtained from the inner profile fitting normalized by the halo mass $M_{\mathrm{NFW}}$ obtained from the outer profile fitting and $M_{\mathrm{NFW}}$, respectively. The orange points with errors show the results from the weak-lensing-only analysis presented in \citet{2026OJAp....961580F}, while the pink and red points with errors show the results from the joint weak and strong lensing analysis using the cored power-law and the broken power-law models, respectively.
The solid line shows the expected relation assuming the Salpeter IMF. A horizontal dotted line shows $\Omega_\mathrm{b}/\Omega_\mathrm{m}$.
{Alt text: The x axis shows the halo mass. The y axis shows the ratio of the stellar mass to the halo mass.}
}\label{fig:shmr} 
\end{figure}

In Figure~\ref{fig:shmr}, we add the newly obtained constraints on the stellar-to-halo mass relation (SHMR) to those presented in \citet{2026OJAp....961580F}, which were derived using the stellar masses constrained by the inner profile fitting and the NFW halo masses constrained by the outer profile fitting. The new constraint is consistent with the result obtained from the weak-lensing-only analysis and provides a tighter constraint on the SHMR. The broken power-law model for the dark matter component yields a tighter constraint, although results for the cored power-law and the broken power-law are consistent with each other. Compared with the relation assuming the Salpeter IMF \citep{2019MNRAS.488.3143B,2025OJAp....8E...8A}, the inferred stellar mass is higher at a given halo mass for the broken power-law case at $\sim 2\sigma$ level. This result favors a bottom-heavy IMF, which is consistent with the result obtained from the weak-lensing-only analysis.

\section{Total density profile}\label{sec:total}

\subsection{Power-law fitting}

We also study the total density profile by combining stacked strong and weak lensing, adopting the fiducial setup of the fitting range described in Section~\ref{sec:fitting}. We consider a simple model of the total density profile assuming a power-law, $\rho(r)\propto r^\gamma$ and constrain the slope $\gamma$. We find $\gamma=-1.96\pm 0.07$, indicating that the total mass profile is fully consistent with a singular isothermal sphere ($\gamma=-2$) within $1\sigma$. The best-fitting $\chi^2$ is 10.4 for the number of degrees of freedom of 20. Since our stacked weak lensing analysis typically probes surface mass densities at $r\lesssim 0.03\,\mathrm{Mpc}/h$ that corresponds to $r\lesssim 10\,r_{\mathrm{e}}$ \citep[see][]{2026OJAp....961580F}, our joint strong and weak lensing analysis suggests that the total mass profile is consistent with a singular isothermal sphere at $r\lesssim 10\,r_{\mathrm{e}}$. This result is in line with previous measurements of total density profiles with strong lensing \citep[e.g.,][]{2006ApJ...649..599K,2013ApJ...777...98S,2018MNRAS.480..431L,2023MNRAS.521.6005E,2024MNRAS.530.1474T,2024A&A...690A.325S,2024SSRv..220...87S}.

\begin{figure}
 \begin{center}
   \includegraphics[width=8.6cm]{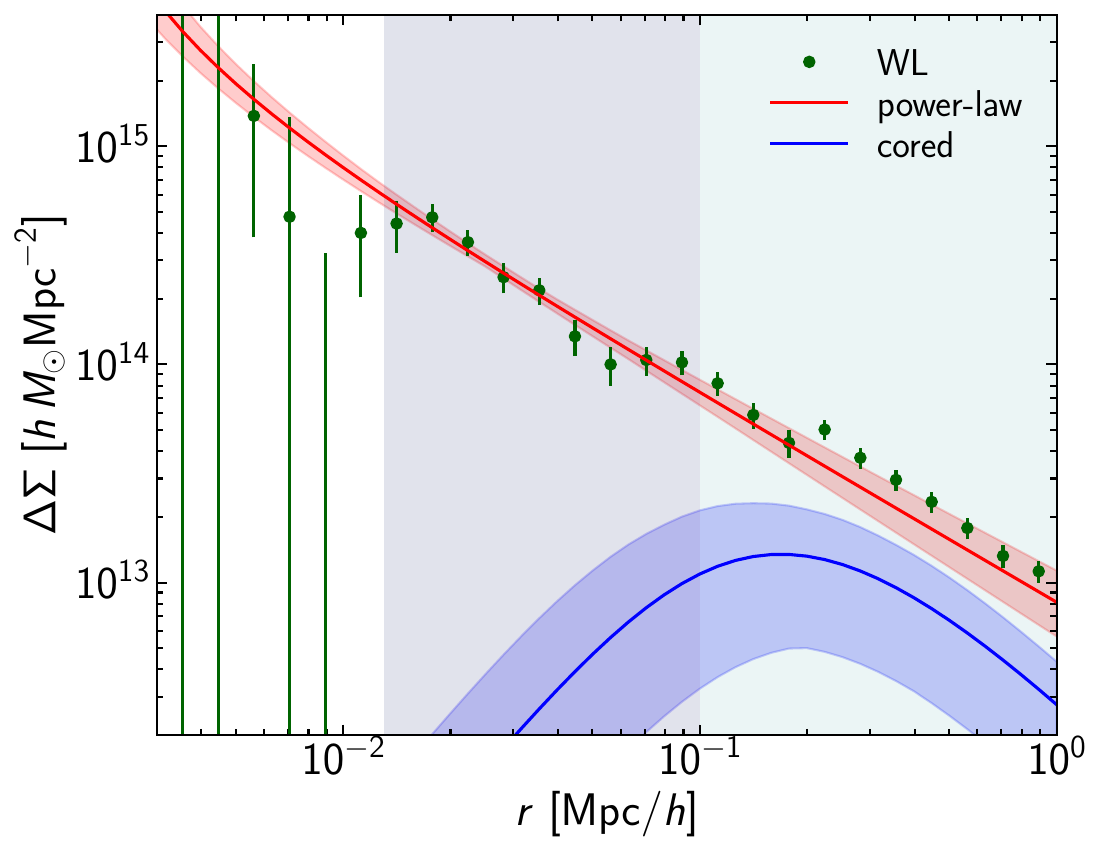} 
 \end{center}
\caption{The differential surface mass density profile $\Delta\Sigma$ measured by stacked weak lensing ({\it green}) is compared with a power-law fitting result with the best-fitting slope of $\gamma=-1.96\pm 0.07$ ({\it red}). It is based on fitting to the strong lensing data as well as the weak lensing data in the inner region indicated by the magenta-shaded region. The best-fitting contribution of the cored component based on fitting to the weak lensing data in the inner plus outer regions indicated by the magenta-shaded region is shown by the blue line.
 {Alt text: The x axis shows the radius from 0.003 to 1 megaparsec per h. The y axis shows the differential surface mass density.}
}\label{fig:wl_pl_msd} 
\end{figure}

Since our weak lensing measurements extend to much outer radii, we can check to what extent the extrapolation of the power-law profile is consistent with those stacked weak lensing data. Figure~\ref{fig:wl_pl_msd} indicates that such extrapolated power-law profile is roughly consistent with the weak lensing data out to $1\,\mathrm{Mpc}/h$. For definiteness, we fit the power-law model using both the strong lensing data and the weak lensing data in the wider range $0.013\,\mathrm{Mpc}/h<r<1\,\mathrm{Mpc}/h$ to find the best-fitting slope of $\gamma=-1.88\pm 0.02$, which is slightly shallower but is consistent with the result above.

\subsection{Mass-sheet degeneracy}

The mass-sheet degeneracy \citep{1985ApJ...289L...1F} is one of the most serious sources of systematic errors in time-delay cosmography \cite[see e.g.,][for a review]{2024SSRv..220...48B}. \citet{2024MNRAS.533..795K} discuss the constraining power of stacked weak lensing for breaking the mass-sheet degeneracy.

To analyze the mass-sheet degeneracy, we adopt a compound mass model used in e.g., \citet{2020A&A...643A.165B}, which is based on a physically motivated profile considered in \citet{2020ApJ...892L..27B}. Specifically, we add the following cored component with its convergence profile
\begin{equation}
 \kappa_{\mathrm{c}}(r)=\frac{1}{1+(r/\tilde{r}_{\mathrm{c}})^2},
 \label{eq:mst_cored}
\end{equation}
where $\tilde{r}_{\mathrm{c}}$ is a core radius, to the power-law profile. As a result, we use the following surface density profile in our analysis 
\begin{equation}
  \Sigma(r)=\lambda_{\mathrm{c}}\frac{3+\gamma}{2}\tilde{A}r^{1+\gamma}+(1-\lambda_{\mathrm{c}})\Sigma_{\mathrm{cr}}\frac{1}{1+(r/\tilde{r}_{\mathrm{c}})^2},
\end{equation}
where $\Sigma_{\mathrm{cr}}$ is computed assuming the source redshift of $z_{\mathrm{s}}=2$, which corresponds to a typical source redshift of lensed quasar systems, and $\tilde{A}$ is a parameter that determines the normalization of the power-law profile. The parameter $\lambda_{\mathrm{c}}$ controls the relative weight of the power-law component and cored component and resembles the mass-sheet transformation parameter when the core radius is much larger than the Einstein radius. 

We constrain $\lambda_{\mathrm{c}}$ from our strong and weak lensing data as follows. Since the power-law slope in the time-delay cosmography analysis is usually determined by strong lensing as well as galaxy kinematics, we consider the situation that the slope $\gamma$ is determined by our fiducial choice of the fitting range of stacked weak lensing signals and hence represents the slope for the inner density profile of galaxies. Since a wide radial range of weak lensing data is crucial to constrain both the amplitude and core radius of the cored component, we fit the two-component model including the cored component using weak lensing data only and use a much wider range of $0.013\,\mathrm{Mpc}/h<r<1\,\mathrm{Mpc}/h$ including the outer part of stacked weak lensing signals. Denoting the former and latter $\chi^2$ as $\chi^2_{\mathrm{SL+WL,in}}$ and $\chi^2_{\mathrm{WL,in+out}}$, respectively, we compute the likelihood as
\begin{equation}
  \mathcal{L}\propto \int d\gamma d\tilde{A}\,\mathcal{L}_{\mathrm{WL,in+out}}(\lambda_{\mathrm{c}},\,\tilde{r}_{\mathrm{c}}|\gamma,\,\tilde{A}),
    \,\mathcal{L}_{\mathrm{SL+WL,in}}(\gamma,\,\tilde{A})
\end{equation}
where $\mathcal{L}_{\mathrm{WL,in+out}}\propto \exp(-\chi^2_{\mathrm{WL,in+out}}/2)$ and $\mathcal{L}_{\mathrm{SL+WL,in}}\propto \exp(-\chi^2_{\mathrm{SL+WL,in}}/2)$, and use $\Delta\chi^2=-2\ln(\mathcal{L}/\mathcal{L}_{\mathrm{max}})$ to derive constraints on $\lambda_{\mathrm{c}}$. The core radius is varied in the range between $10^{-1.5}\,\mathrm{Mpc}/h$ and $10^{-0.5}\,\mathrm{Mpc}/h$ to ensure that weak lensing data covers the inside and outside the core radius and is marginalized over. We find $\lambda_{\mathrm{c}}=0.978\pm 0.010$. The resulting contribution of the cored component from this analysis is shown in Figure~\ref{fig:wl_pl_msd}. Our result indicates the modest but non-zero impact on the mass-sheet degeneracy for analyses assuming a power-law profile for the total density profile of lensing galaxies, which is consistent with previous work \citep[e.g.,][]{2025MNRAS.538.2375L,2025MNRAS.541....1S}.

{We caution that the tight constraint on the mass-sheet transformation parameter $\lambda_{\mathrm{c}}$ derived above is for a specific case assuming a single cored component, and the constraint can be relaxed once more general mass-sheet transformation is considered. As a specific example, following \citet{2024MNRAS.533..795K}, we consider a superposition of three cored profiles defined by Equation~\eqref{eq:mst_cored} with fixed core radii of $\tilde{r}_{\mathrm{c}1}=0.05\,\mathrm{Mpc}/h$, $\tilde{r}_{\mathrm{c}2}=0.1\,\mathrm{Mpc}/h$, and $\tilde{r}_{\mathrm{c}3}=0.2\,\mathrm{Mpc}/h$, and fit the three normalization parameters $\lambda_{\mathrm{c}1}$, $\lambda_{\mathrm{c}2}$, and $\lambda_{\mathrm{c}3}$. We obtain $\lambda_{\mathrm{c}}=1-\sum_i (1-\lambda_{\mathrm{c}i})=0.99\pm 0.03$, which is consistent with the result for the case of one cored component, but with a larger error. }

\section{Conclusion}\label{sec:conlusion}

We have studied the stellar and dark matter distributions of massive ($M_\star \sim 10^{11.6}M_\odot$) early-type galaxies at $0.4<z<0.6$ utilizing both strong and weak lensing measurements. We have combined stacked weak lensing measurements down to small radii, which is obtained from the HSC-SSP three-year shape catalog, presented in \citet{2026OJAp....961580F} with stacked strong lensing measurements of enclosed projected masses from 13 strong lens systems selected from various surveys. We have corrected for the strong-lensing selection bias using mock strong lens catalogs \citep{2025OJAp....8E...8A}. We have found that adding strong lensing constraints indeed improves constraints on the stellar and dark matter distributions. The central dark matter density profile is found to be consistent with the NFW profile. Our direct stellar mass measurements with the joint strong and weak lensing analysis favor a bottom-heavy stellar IMF. 

We have also studied the total density profile over a wide range in radii. We have found that the central total density profile is consistent with a power-law profile with the slope of $\gamma=-1.96\pm 0.07$. The extrapolation of the best-fit power-law profile explains the weak lensing data reasonably well out to $1\,\mathrm{Mpc}/h$. We have used our strong and weak lensing data to constrain the mass-sheet transformation parameter $\lambda_{\mathrm{c}}$, which quantify the slight deviation from the power-law, as $\lambda_{\mathrm{c}}=0.978\pm0.010$, indicating the non-zero mass-sheet when assuming the power-law profile. {The constraint is relaxed to $\lambda_{\mathrm{c}}=0.99\pm0.03$ when we consider a superposition of three cored components with fixed core radii.} 

The methodology developed in this paper paves the way to the statistical analysis of density profiles of galaxies with stacked strong and weak gravitational lensing in e.g., the Euclid Wide survey \citep{2025A&A...697A...1E}. {On the other hand, one of the major 
limitations of the current analysis presented in the paper is 
the heterogeneous nature of the strong-lens sample, which makes an 
accurate characterization of the strong-lensing selection bias difficult.}
In the analysis of such future survey data, a well-defined sample of strong lens systems should be used to quantify the strong-lensing selection bias more cleanly. The large sample of strong lenses available in future surveys will also enable one to constrain the stellar and dark matter distributions of galaxies over a wide range in stellar masses and redshifts.

\begin{ack}
{We thank an anonymous referee for useful comments.}
This work was supported by JSPS KAKENHI Grant Numbers JP25H00662, JP25H00672, JP22K21349, JP24K00684.

The Hyper Suprime-Cam (HSC) collaboration includes the astronomical communities of Japan and Taiwan, and Princeton University. The HSC instrumentation and software were developed by the National Astronomical Observatory of Japan (NAOJ), the Kavli Institute for the Physics and Mathematics of the Universe (Kavli IPMU), the University of Tokyo, the High Energy Accelerator Research Organization (KEK), the Academia Sinica Institute for Astronomy and Astrophysics in Taiwan (ASIAA), and Princeton University. Funding was contributed by the FIRST program from the Japanese Cabinet Office, the Ministry of Education, Culture, Sports, Science and Technology (MEXT), the Japan Society for the Promotion of Science (JSPS), Japan Science and Technology Agency (JST), the Toray Science Foundation, NAOJ, Kavli IPMU, KEK, ASIAA, and Princeton University. 

This paper makes use of software developed for the Large Synoptic Survey Telescope. We thank the LSST Project for making their code available as free software at  http://dm.lsst.org

This paper is based on data collected at the Subaru Telescope and retrieved from the HSC data archive system, which is operated by the Subaru Telescope and Astronomy Data Center (ADC) at National Astronomical Observatory of Japan. Data analysis was in part carried out with the cooperation of Center for Computational Astrophysics (CfCA), National Astronomical Observatory of Japan. The Subaru Telescope is honored and grateful for the opportunity of observing the Universe from Maunakea, which has the cultural, historical and natural significance in Hawaii. 
\end{ack}


\section*{Data availability} 
 The data underlying this article will be made available at {\tt https://hsc.mtk.nao.ac.jp/ssp/}.

 \bibliographystyle{apj}
 \bibliography{refs}

\end{document}